\documentclass[prl,showpacs,twocolumn]{revtex4}
\usepackage{graphicx}
\usepackage{amssymb}
\usepackage{bm}

\newcommand{\be}{\begin{eqnarray}}
\newcommand{\ee}{\end{eqnarray}}
\newcommand{\nn}{\nonumber}
\newcommand{\bn}{\begin{enumerate}}
\newcommand{\en}{\end{enumerate}}

\def\IC{\mathbb{C}}

\def\IR{\mathbb{R}}
\def\IZ{\mathbb{Z}}

\def\CN{{\cal N}}

\def\b{\beta}

\def\d{\delta}
\def\e{\epsilon}

\def\t{\tau}

\def\D{\Delta}

\def\S{\Sigma}

\def\O{\Omega}

\def\half{\frac{1}{2}}

\def\goto{\rightarrow}

\begin{document}

\title{Superconformal field theories from crystal lattices}
\author{Sangmin \surname{Lee}}
\affiliation{Department of Physics and Astronomy, 
Seoul National University, Seoul 151-747, Korea}

\begin{abstract}
We propose a brane configuration for the (2+1)d, $\CN=2$ 
superconformal theories (CFT$_3$) arising from M2-branes 
probing toric Calabi-Yau 4-fold cones, 
using a T-duality transformation of M-theory. 
We obtain intersections of M5-branes on a three-torus 
which form a 3d bipartite crystal lattice in a way similar to 
the 2d dimer models for CFT$_4$. 
The fundamental fields of the CFT$_3$ are M2-brane discs localized around 
the intersections, and the super-potential terms are identified with the atoms 
of the crystal.  
The model correctly reproduces the complete BPS spectrum of 
mesons and baryons.

\end{abstract}

\pacs{11.25.-w, 04.50.+h, 04.20.Jb}

\maketitle


Extending the AdS/CFT correspondence \cite{adscft} to theories 
with less than maximal supersymmetry has been an important subject 
from early years \cite{kw,ach,mp}. 
For four dimensional theories (CFT$_4$), remarkable progress has been 
made recently. The discovery of new AdS solutions \cite{gmsw1, clpp} 
was accompanied by identification of the corresponding quiver gauge theories 
\cite{ms1,bfhms,bk,bfz}. 
It was soon realized that theories arising from D3-branes probing a toric 
Calabi-Yau 3-fold (CY$_3$) singularities share common simplifying features. 
All essential properties of the CFT$_4$ are encoded in a certain periodic 
bipartite graph in two dimensions \cite{hken,fhkvw,fhm,hv,fhkv,cep} 
often called a dimer model.
The algorithm to go from a given toric CY to a dimer graph and vice versa 
has been firmly established \cite{hv,fhkv}. 

Much less is known about three-dimensional theories 
arising from M2-branes near CY$_4$ singularities. 
There is no qualitative difference on the geometry side; 
toric descriptions are known \cite{ms1} for the new supergravity solutions 
\cite{gmsw2} and many results in toric geometry 
are insensitive to the dimension of the CY \cite{msy1}. 
In contrast, conventional field theoretic tools are of little use 
because the CFT$_3$ is necessarily strongly coupled.
Even the simplest case of M2-branes in flat space remains elusive. 

In this paper, we build upon the intuition from the 2d dimer model for CFT$_4$ 
to show that each toric CFT$_3$ can be related to a 3d crystal lattice.  
The construction uses T-duality of M-theory on a three-torus 
just as a T-duality of IIB string theory is used to 
derive the dimer model \cite{fhkvw}.
Our model maps the CFT$_3$ to the world-volume theory of M5-branes 
intersecting along the bonds of the crystal lattice. 
Although the M5-brane theory is as poorly 
understood as the CFT$_3$ itself, we will be able to use our model to 
make some quantitative statements.

First, we show that the fundamental excitations of the CFT$_3$ are 
M2-brane discs localized around M5-brane intersections. 
Second, we show how to evaluate various global charges 
of the baryons and mesons and explain how they are related. 
Finally, we identify the super-potential terms which yield the F-term conditions.
We use them to construct the spectrum of BPS mesons, 
which agrees perfectly with the corresponding result from the geometry.

We begin with a stack of $N$ M2-branes near the tip of a CY$_4$ cone $X$. 
In the near horizon limit, the $\IR^{1,2}$ world-volume directions 
of the M2-branes 
and the radial direction of $X$ merge to form an AdS$_4$ 
and the base (unit radius section) of $X$ becomes 
the internal 7-manifold $Y$. We say $X$ is the cone over $Y$, or $X=C(Y)$. 
The cone $X$ being K\"ahler and Ricci-flat 
is equivalent to the base $Y$ being Sasakian and Einstein, respectively. 

Our construction assumes that $X$ is toric, so we first recall 
some relevant aspects of toric geometry \cite{msy1}. 
The toric diagram is a convex polyhedron composed of a set of lattice points 
$\{v_I^i\} \in \IZ^4$ ($I=1, \cdots, d \ge 4$). 
The CY condition requires that the $v_I$ be on the same $\IZ^3$ subspace. 
It is customary to choose a basis to set $v_I^4=1$ for all $I$. 

The toric diagram defines a solid cone 
$\D \equiv \{ y_i \in \IR^4 ; (v_I \cdot y) \ge 0 \mbox{ for all }I\}$. 
We call the boundary components 
$S^I \equiv \D \cap \{v_I\cdot y = 0\}$ the 3-fans. 
Two 3-fans meet at a 2-fan and several 2-fans join at a 1-fan. 
These fans are graph dual to the original toric diagram in 
the sense that each vertex $v_I$ is associated to a 3-fan, 
each edge connecting two neighboring vertices, $w_{IJ} = v_I -v_J$, 
corresponds to a 2-fan, etc. 
The cone $X$ is a $T^4$ fibration over $\D$. 
The fiber is aligned with the base in such a way that it shrinks to $T^n$ on 
the $n$-fans.

\begin{figure}[htb]
\includegraphics[width=4.0cm]{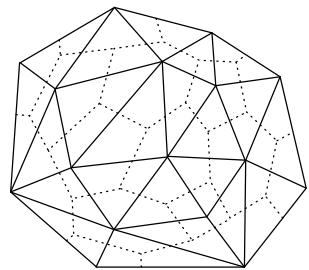}
\caption{A toric diagram (solid line) is
a convex polyhedron with integer-valued vertices in $\IR^3$. 
Its graph dual (dashed line) gives the fan diagram.} \label{partition}
\end{figure}

The moduli space of K\"ahler metrics on $X$ is parameterized by the Reeb vector 
$b^i \in \IR^4$, which defines the base of the cone by 
$Y = X \cap\{b\cdot y = 1/2\}$. 
In the basis mentioned above, the CY condition fixes $b^4=4$. 
The volume of $Y$ as an explicit function of $v_I$ and $b$ is known. 
The Ricci-flat metric is obtained by minimizing the volume with respect to 
$b^i$ with the range of $b^i/4$ being precisely the toric diagram. 

For later purposes, we recall that a baryon in the CFT$_3$ is mapped 
via AdS/CFT to an M5-brane wrapping a supersymmetric 5-cycle of $Y$ 
\cite{bhk}.
Each 3-fan $S^I$ dual to a vertex $v_I$ defines a 5-cycle 
which is the $T^3$ fibration over $S^I \cap \{b\cdot y = 1/2\}$. 
We abuse the notation a bit and use $S^I$ to denote either 
the 3-fan or the 5-cycle depending on the context. 
The baryons are charged under the global symmetries of the theory. 
The four $U(1)$ isometries $F_i$ of $X$ are called flavor symmetries 
and there are also $d-4$ baryonic symmetries $Q_a$, where $d$ is the 
number of vertices of the toric diagram. 
The charges of the the baryons 
$F_i[S^I]  \equiv N F_i^I$, $Q_a[S^I] \equiv N Q_a^I$
satisfy the toric relations $v_I^i F^I_j = \d^i_j$ and $v_I^i Q_a^I = 0$ \cite{LR}.
The $U(1)_R$ charge is the linear combination of the flavor charges 
$R = \half b^i F_i$, where $b^i$ is the Reeb vector. 
The toric relations and the CY condition ($b^4=4$) implies that $\sum_I R^I = 2$.

Our crystal model follows from a T-duality of M-theory. 
It was inspired by a similar derivation of the 2d dimer model 
using T-duality of IIB string theory \cite{hken}. 
We take the T-duality transformation along $T^3\subset T^4$ aligned with the
$y_{1,2,3}$ coordinates.
By T-duality, we mean the element $t$ in the $SL(2,\IZ)\times SL(3,\IZ)$ duality 
group which acts as $t: \tau \equiv C_{(3)} + i \sqrt{g_{T^3}} \goto -1/\tau$.
The stack of $N$ M2-branes turns into a stack of $N$ M5-branes wrapping 
the dual $T^3$. 
We call them the $T$-branes. 
The degenerating circle fibers turn into another M5-brane 
extended along the (2+1)d world-volume and a non-trivial 3-manifold 
$\S$ in $\IR^3 \times T^3$. We call it the $\S$-brane. 
The result is summarized in Table I.

\begin{table}[htb]
\label{brane3}
$$
\begin{array}{l|ccc|cccccc|cc}
\hline
           & 0 & 1 & 2 & 3 & 4 & 5 & 6 & 7 & 8 & 9 & 11\\
\hline
\mbox{M5} & \circ & \circ & \circ & & & & \circ & \circ & \circ \\
\mbox{M5} & \circ & \circ & \circ &  \multicolumn{6}{c|}{\Sigma} & & \\
\hline
\end{array}
$$
\caption{The brane configuration for the CFT$_3$.  
The special Lagrangian 3-manifold $\Sigma$ is locally a product of 
a surface in $\IR^3$(345) and a curve in $T^3$(678).}
\end{table}

The duality chain shows that $\S$ is locally a product of a surface in $\IR^3$ and 
a curve in $T^3$. 
The geometry of $\S$ can be read off from the toric data. 
In particular, the projection of $\S$ onto $\IR^3$ is a thickened 
version of the collection of the 2-fans. This is similar 
to the so called amoeba projection used to analyze the 2d dimer model. 
Note that a 2-fan dual to an edge $w_{IJ}=(p,q,r)$ is locally a plane 
orthogonal to $w_{IJ}$ in $\IR^3$. 
Away from the intersections with other 2-fans, 
$\S$ is locally the product of this plane and the 
$(p,q,r)$ 1-cycle in the $T^3$.  
The correlation of the orientation ensures that $\S$ is locally special Lagrangian, 
i.e., it is calibrated by $\mbox{Im}\O$, 
where $\O$ is the holomorphic three-form 
$\O = (dx^3+idx^6)\wedge(dx^4+idx^7)\wedge(dx^5+idx^8)$, 
and the pull-back of the K\"ahler form onto $\S$ vanishes. 
This is in accordance with the $\CN=2$ supersymmetry on the (2+1)d world-volume.

In analogy with the D5/NS5-brane description of the dimer model, 
we expect that the content of the CFT$_3$ should be encoded 
in the intersection locus between 
the $T$-branes and the $\S$-brane projected onto the $T^3$. 
The result is a graph in the $T^3$ which we call the crystal lattice.  
As in the 2d dimer model, the graph consists of edges and 
vertices which we call {\em bonds} and {\em atoms} 
to distinguish them from similar objects in the toric diagram. 

To obtain a bond, we need a collection of 2-fans to form 
a closed region in $\IR^3$ containing one or more vertex $v_I$.  
When the $(p,q,r)$-cycles of the $T^3$ associated to the 2-fans 
join along a line segment, a bond is formed. 
Note that while the 2d dimer model allows only a pair of 1-cycles to merge 
and form an edge, in our crystal model three or more 1-cycles can merge 
along a bond. 

\begin{figure}[htb]
\includegraphics[width=4.0cm]{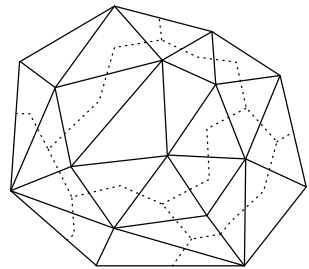}
\caption{A partition of the toric diagram.} \label{partition2}
\end{figure}

To see how several bonds meet at an atom, 
consider a partition of the toric diagram (Figure 2). 
Each closed region gives a bond as explained above. 
The bonds can meet at an atom only if 
the regions add up to cover the entire toric diagram \cite{LL}.

Figures 3 and 4 illustrates the rules with the two simplest examples, 
namely, $\IC^4$ and 
the cone over the homogeneous space $Q(1,1,1)$. 
The latter example has been studied 
in the AdS/CFT context in Refs. \cite{ahn1, oh, fab, ahn2, glmw}. 
Note that the crystal lattice inherits the discrete symmetry of the toric diagram. 
The tetrahedral symmetry of $\IC^4$ is reflected in its 
lattice which has the familiar zincblende (GaAs) structure. 
The $C(Q(1,1,1))$ cone and its lattice which has the NaCl structure 
share the octahedral symmetry. 
More examples and details of the general algorithm to determine the crystal structure will be reported in Ref. \cite{LL}.

\begin{figure}[htb]
\includegraphics[width=8.0cm]{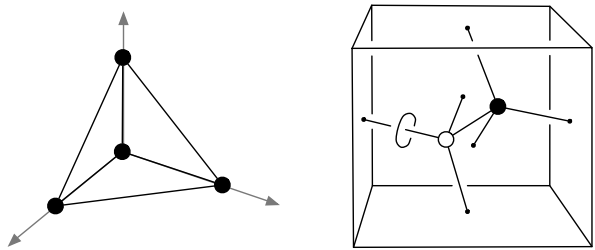}
\caption{The crystal for $\IC^4$ has the zincblende structure.} \label{C4}
\end{figure}

\begin{figure}[htb]
\includegraphics[width=8.0cm]{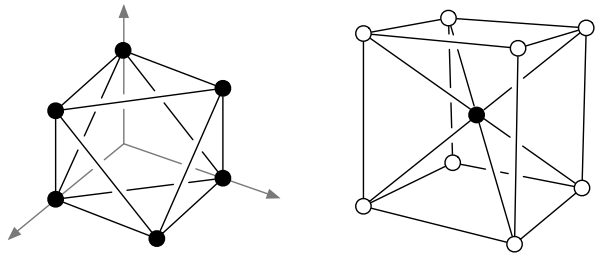}
\caption{The crystal for $C(Q(1,1,1))$ has the NaCl structure.} \label{Q111}
\end{figure}


We now proceed to study the physical properties of the CFT$_3$ 
using our crystal model. First of all, we have to figure out 
what the the fundamental excitations are and where in the crystal 
they live.
Recall that we have a known description of the baryons on the geometry side. 
Our strategy is to bring the baryon to the crystal picture 
and try to deform it smoothly into a collection of $N$ fundamentals, 
thereby identifying the latter. 
A similar process in the 2d dimer model was discussed in Ref. \cite{ima}.

The T-duality transformation turns the baryon M5-branes 
in the original picture into M2-branes. 
Since the projection of the $\S$-brane onto $\IR^3$ realizes 
the toric fans, the baryon M2-brane corresponding to a 3-fan 
must have a boundary on the 2-fan walls surrounding it. 
Topologically, the M2-baryon is a disc. 
We can shrink the baryon until we approach 
the intersection between the T-branes and the $\S$ brane. 
The intersection is a bond in the crystal formed by the 
1-cycles paired with the 2-fans. The boundary of the M2-baryon 
is now a small circle on the $\S$ brane localized around the intersection. 
We now show that the baryon can be smoothly deformed 
into a collection of $N$ M2-branes with its boundary on the T-branes 
localized around the intersection. 

Consider a supersymmetric configuration of two M5-branes 
intersecting along a (3+1)-dimensional space.
Keeping the mutually transverse directions only, 
the intersection can be described locally by a holomorphic equation 
$zw=0$ with $(z,w)\in \IC^2$. 
Suppose we have an M2-brane disc with boundary along a circle around the 
origin in the $z$-plane. Infinitesimal deformation of the intersection into 
$zw=\e$ shows that the circle on the $z$-plane can be 
smoothly deformed into another circle on the $w$-plane. 
The argument is easily generalized to multiple branes. 
An M5-brane on the $z$-plane intersecting with $N$ M5-branes 
on the $w$-plane is described by $z^N w = 0$. 
The same argument as above shows that an M2-disc on the $z$-plane 
now turns into a stack of $N$ M2-discs on the $w$-plane. 
Thus we find that the fundamental excitations of the CFT$_3$ are the 
M2-discs whose boundary encircles the bonds of the crystal lattice. 
This fundamental M2-disc is the analog of the bi-fundamental  
field localized around the D5/NS5 intersection in the 2d dimer model.

The fundamental M2-discs explain why it is difficult, if not impossible, 
to write down a field theory Lagrangian for the CFT$_3$ 
(See Refs. \cite{oh, ahn1, fab} for previous attempts). 
Our model suggests that in order to write write down the field theory, 
we first have to understand the quantum theory of M2-discs on the M5-brane world-volume.
The CFT$_3$ is not likely to be an ordinary any gauge theory.  
Recall that each face of the 2d dimer graph gives an $SU(N)$ gauge group 
in the quiver gauge theory, and the matter fields are 
bi-fundamentals connecting adjacent gauge groups. 
In contrast, all non-localized M2-discs in our crystal model are equivalent, 
and it is not clear whether they represent gauge fields.

The transition from a baryon to fundamentals also gives 
an easy way to find the global charges of the fundamentals, 
as the fundamentals simply inherit the charges of the baryons.  
If a fundamental M2-disc originates from a region of 
a partition shown in Figure 2, its global charges are $1/N$ times 
the sums of charges ($F^I_i$, $Q_a^I$) of the 3-fans $S^I$ inside the region. 

The M2 discs localized along the bonds meeting at an atom 
can form a spherical membrane. In analogy with 
the 2d dimer model, we identify it as a super-potential term of the CFT$_3$.  
On general grounds, we expect that the super-potential terms 
have $R$-charge two and vanishing flavor charges. 
It is indeed true because (i) every atom corresponds to a partition covering the entire toric diagram (ii) the sum of charges over all 3-fans 
is either two ($R$-charge) or zero (flavor charges) due to the toric relations.
When two atoms are connected by a bond, the M2-disc 
contribute to the two super-potential terms with opposite orientation. 
If we paint the atoms according to the orientation, then 
only atoms with opposite colors can be connected.
In other words, the crystal is bipartite. 

The super-potential terms yield the F-term equivalence relations among the 
elements of the chiral ring. 
As in the 2d dimer models, the relations implies that the 
sum of terms corresponding to any two connected vertices 
are F-term equivalent to zero. 

Chiral mesons in CFT$_3$ correspond to algebraic (holomorphic polynomial) functions on the CY$_4$ cone $X$. They are labeled by integer 
points $m=(m_1,m_2,m_3,m_4)$ in the solid cone $\D$, 
which are the momentum quantum numbers 
along the $T^4$ fiber of $X$.
It follows immediately that the flavor charges of a meson $m$ are $F_i(m) = m_i$. 
The $R$-charge of the meson is then given by \cite{msy2, gmsy, zaffnew}
\be
R(m) 
= \frac{1}{2} b^i F_i(m) = \frac{1}{2} (m \cdot b).
\ee

Under the T-duality transformation, the mesons are transformed into 
closed M2-branes. 
The first three components of $m$ define the homology 2-cycle of the 
M2-meson along the $T^3$. It is instructive to confirm this fact 
in the crystal model.  
When an M2-meson wraps a 2-cycle, its charge can be computed 
from its intersections with the bonds in the crystal, 
as the meson is a bound state of the fundamental M2-discs. 
Since the bonds are made of $(p,q,r)$-cycles corresponding to 
the edges $w_{IJ}$ of the toric diagram, we may equally well compute the  
charges from the intersection of the meson with the $(p,q,r)$-cycles.
Following Ref. \cite{buttinew}, we use the fact that $\sum_I F_i^I = 0$ 
($i \ne 4$) to assign variables $F_i^{IJ}=-F_i^{JI}$ to the edges $w_{IJ}$, 
such that 
$F_i^I = \sum_{J} F_i^{IJ}$, where the sum runs over the neighboring vertices. 
Then we find that the charge of the M2-meson is indeed given by 
\be
F_i (m) &=& \sum_{(IJ)} (m\cdot w_{IJ}) F_i^{IJ} 
= \sum_{(IJ)} m \cdot(v_I-v_J) F_i^{IJ} 
\nn \\ 
&=& \sum_I (m \cdot v_I) F^I_i  = m_i . 
\ee
The last component of $m$, contributing $2m_4$ to the $R$-charge, 
measures how many times the M2-meson wraps a super-potential term. 
The F-term condition discussed above 
guarantees that all the M2-meson with the same value of $m$ 
are F-term equivalent, regardless of the precise way they wrap the 2-cycle. 
So there is a unique meson for each value of $m$. 
This perfectly matches the known results on the BPS spectrum of chiral mesons from the geometry side \cite{msy2, gmsy, zaffnew, counting}. 

Some remarks are in order. Both for the 2d dimer and 3d crystal models, 
the derivation based on T-duality connect objects on the geometry side 
such as baryons and mesons with their counterparts on the CFT in a smooth 
manner. It realizes in a concrete way the common lore that the AdS/CFT correspondence is a open/closed duality. We also note that the derivation of the 
$\b$-deformed geometry of the toric CY cones discussed in Ref. \cite{lunin} 
tacitly assumed a model for CFT$_3$ of the type we presented here. 

There are several issues that deserve further investigation.  
It is well-known that the 2d dimer model for a given CY$_3$ is not unique 
and different models are related by Seiberg duality. It would be interesting 
to see whether similar phenomena occurs in the crystal model. 
The analysis in this paper was restricted to the topological aspects of the model. 
It would be important to extend it to include dynamical issues 
such as non-BPS meson spectrum, moduli space of vacua, 
marginal deformations other than the $\b$-deformation \cite{lunin, ahn2, glmw},  and volume minimization \cite{msy1} and $\t_{RR}$-minimization \cite{trr}.
Finally, it would be interesting to look for a brane configuration 
for the $\CN=3$ CFT$_3$ \cite{billo, ly}.

\begin{acknowledgments}
We thank Surjay Lee, Jaemo Park and Soo-Jong Rey for useful discussions 
and comments on the manuscript.
We are especially grateful to Amihay Hanany for a series of lectures 
at the APCTP/CQueST/KIAS Winter School on String 
Theory in February 2006.
This work was supported by the Faculty Grant of Seoul National University and 
the KOSEF Basic Research Program, grant No. R01-2006-000-10965-0.
\end{acknowledgments}

\end{document}